\documentclass[twocolumn]{revtex4}
\usepackage{bm}
\usepackage{graphicx,amssymb,amsmath} 

\newcommand {\be}  {\begin{equation}}
\newcommand {\ee}  {\end{equation}} 
\newcommand {\bea} {\begin{eqnarray}}
\newcommand {\eea} {\end{eqnarray}}
 
\begin{document}

\title{Quantum Statistical Physics of  Glasses at Low Temperatures}
\author{J. van Baardewijk and R. K\"uhn.}
\address{Department of Mathematics,King's College University-London,Strand,London,WC2R.2LS,UK.}
\date{\today}

\begin{abstract}
{\scriptsize We present a  quantum statistical analysis of a microscopic mean-field model of structural glasses at low temperatures. The model can be thought of as arising from a random Born von Karman expansion of the full interaction potential. The problem is reduced to a single-site theory  formulated in terms of an imaginary-time path integral using replicas to deal with  the disorder. We study the physical properties of the system in thermodynamic equilibrium and develop both perturbative and non-perturbative methods to solve the model. The perturbation theory is formulated as a loop expansion in terms of two-particle irreducible diagrams, and is carried to three-loop order in the effective action. The non-perturbative description is investigated in two ways, (i) using a static approximation, and (ii) via Quantum Monte Carlo simulations. Results for the Matsubara correlations at two-loop order perturbation theory are in good agreement with those of the Quantum Monte Carlo simulations. Characteristic low-temperature anomalies of the specific heat are reproduced, both in the non-perturbative static approximation, and from a three-loop perturbative evaluation of the free energy. In the latter case the result so far relies on using Matsubara correlations at two-loop order in the three-loop expressions for the free energy, as self-consistent Matsubara correlations at three-loop order are still unavailable. We propose to justify this by the good agreement of two-loop Matsubara correlations with those obtained non-perturbatively via Quantum Monte Carlo simulations.
}\end{abstract}
\maketitle

\section{Introduction}
Glasses are known to exhibit distinctive low-temperature properties that differ substantially from those of crystalline solids and are referred to as glassy low-temperature anomalies.
For instance, at  low temperatures the specific heat and thermal conductivity in crystals show a familiar $T^3$-dependence. In glasses the specific heat is found to  increase approximately linearly with the temperature  at $T<1\, K$, while the thermal conductivity increases approximately as $T^2$ in this low temperature range \cite{ZellPohl71}.
At higher temperatures between 1 and 20$\,K$ the thermal conductivity is approximately constant, while the specific heat $C$ shows a peak when displayed as  $C/T^3$, usually referred to as the Bose-peak.
In phenomenological models such as the Standard Tunnelling Model \cite{And+72,Phil72} and the soft potential model \cite{Karp+83,Buch+92} one postulates that a broad spectrum of tunnelling centres is responsible for the properties at $T<1\, K$.

Surprisingly, the glassy anomalies   show a noticeable degree of universality at $T<1\, K$, whereas between approximately 1 and 20 K these depend more on the specific materials \cite{ZellPohl71}.
There are currently two main contending theories to explain this fact. Following ideas of Yu and Leggett \cite{YuLegg88}, it has been suggested as resulting from a collective effect due to interactions between the tunnelling excitations \cite{BurKag96a}.
Alternatively, it is thought to be a property of the potential energy landscape  created by glassy freezing at high temperatures. This also defines a phenomenon of  collective origin but involves no quantum effects \cite{KuHo97}. Universality in the second interpretation is understood as a result of separation of the energy scales involved in glassy freezing on the one hand side, and those relevant for the low-temperature phenomena on the other hand side \cite{KuHo97,Kuehn03}.
Whereas the existence of tunnelling centres is part of the initial assumptions in \cite{BurKag96a}, these are shown to arise naturally as a result of \emph{microscopic} interactions in the model glasses studied in \cite{KuHo97,Kuehn03}, and do indeed give rise to the characteristic low-temperature anomalies. The work in \cite{KuHo97,Kuehn03} is perhaps appropriately characterised as a strong coupling approach to glassy low-temperature physics. 
A complementary weak coupling approach \cite{Gurevich+03,Parshin+07} to the same phenomena takes weak residual interactions between a set of quasi-local collective modes as starting point and describes low-temperature anomalies in terms of a vibrational instability occuring in systems of this type.

The analysis in \cite{KuHo97,Kuehn03} is  still semi-classical   in the sense that it is based on an analysis of quantum effects in a glassy potential energy landscape whose properties were determined via classical statistical mechanics. The aim of the present paper is to overcome this deficit and study the system in a full quantum statistical formulation right from the outset.  Focus will be here  on the translationally invariant model proposed in  \cite{Kuehn03}.
 
We shall proceed along the lines of general methods developed for quantum spin-glasses. In particular, we apply the Matsubara formalism to construct an imaginary-time path integral representation of the partition function and the replica-method to deal with the disorder. The sites are decoupled by introducing order parameters for which the functional integral is evaluated by the method of steepest descent. The result is an effective single-site theory and a set of 
functional self-consistency relations for the order parameters. These methods are similar to those  used for models studying spin-glass transitions  in  quantum spin-glasses. Examples are the SK-model of spin-glasses generalised to quantum spins \cite{BrayMoore80} and the quantum spherical $p$-spin glass model \cite{NieuwRit98,Cugl+01}. Here we shall not concern ourselves with the glass transition but concentrate on evaluating the physical properties at low temperatures and  in particular the specific heat anomaly in the $1\,K$ region. 

To solve the effective single-site theory we first apply a perturbative method in terms of two-particle irreducible (2PI) diagrams, which is based on an expansion in powers of the full interacting correlation functions. This amounts to summing infinite classes of diagrams and can therefore also capture effects of a non-perturbative nature. 
After this  we develop a non-perturbative theory proper. 
The result is a set of functional self-consistency equations for the order parameters which we first treat with Quantum Monte Carlo simulations.

 Following this we construct a \emph{solvable} version of the non-perturbative theory, using a simple approximation known from  quantum spin-glass theory as the \emph{static} approximation. This  scheme  was first introduced as a variational Ansatz in \cite{BrayMoore80} where the time-dependent order parameter was approximated by a time-independent constant. 

Given the complications of dealing with quantum fluctuations in this model, we presently restrict the analysis of both perturbative and non-perturbative theory to the replica symmetric approximation. In support of this we mention that the effects  of replica symmetry breaking on the low-temperature anomalies were found to be small at the semi-classical level \cite{KuHo97,Kuehn03}. \\

This paper is organised as follows. In section \ref{sec:Model} we summarise the main ingredients of the proposed glass model. In section  \ref{sec:Partition} we give the many-particle partition function represented by an imaginary-time path integral and introduce replicas to handle the disorder averaging.  Section \ref{sec:SS}  gives an  account of the effective single-site formulation, deriving the effective action and the functional self-consistency relations for the order parameters. 
Then  in section \ref{sec:pert} we treat perturbative and in section \ref{sec:non-pert} non-perturbative solution methods. The numerical results for the order parameters and the specific heat are discussed in section \ref{sec:results}.  Finally, the conclusions are drawn in section \ref{sec:conclusions}.

\section{The glass model}\label{sec:Model}
Starting point is the microscopic model for a glass-like system at low temperatures, proposed in \cite{Kuehn03}. 
To summarise its main ingredients, consider a system of $N$ degrees of freedom (called particles)  with the following model-Hamiltonian
\be\label{eq:H}
H=T+V=\sum_{i=1}^N\frac{p_i^2}{2m_i}+V(\bm{u}),
\ee
where the $p_i$ denote the momenta  and  the $m_i$ the masses, which for simplicity are taken equal for each particle.  
The variables $\bm{u}=(u_1,..,u_N)$ represent  coordinate deviations from pre-assigned reference positions. Since glassy low-temperature physics is universal, the interaction potential  $V(\bm{u})$ need not contain details of the specific atoms and their specific interactions  
and is taken as simple as possible, yet containing enough detail to reproduce glassy low-temperature physics. 
The minimum requirement is that it should respect global translation invariance and have elements of randomness and frustration.
There are two possibilities to make analytic progress: use a mean-field approximation for a microscopically `semi-realistic' model or alternatively formulate a model for which such a mean-field approximation would be exact. The latter is the approach we have taken here, a justification of which  we believe is provided by the results. 
The absence of phonons is of course one of the unavoidable consequences of adopting a mean-field approximation.

Following \cite{Kuehn03}, the potential function is taken to represent the first terms of a Born von Karman expansion of the full interaction energy about the reference positions.
A further requirement of  global $Z_2$-symmetry ($\bm{u}\leftrightarrow -\bm{u}$) excludes the odd orders in this expansion.  Matters are further simplified  by taking the $u_i$ to be scalar, resulting in
\be\label{eq:Vu}
V(\bm{u})=\sum_{i<j}^N\Big[\,\frac{1}{2}\,J_{ij}\,(u_i-u_j)^{2}+\frac{g}{N}\,(u_i-u_j)^4\,\Big].
\ee
The glassy properties are represented  at the quadratic level in (\ref{eq:Vu}), defining a random-interaction term with  random interaction strengths $J_{ij}$. The quartic term (taken to be non-random) is necessary in order to stabilise the system as a whole, and so $g>0$. 
The parameters $J_{ij}$ are quenched and taken independent  with equal Gaussian distribution ${\cal N}(0,J^2/N)$ for each combination  ($i,j$). The  $1/N$ scaling of the variance and the quartic interaction term in (\ref{eq:Vu}) ensures that the thermodynamic energy is proportional to $N$. 
The construction as presented here allows the system to be analysed within replica mean-field theory, similar to that of the SK-model for spin-glasses \cite{SK75}, its generelization to quantum spin-glasses \cite{BrayMoore80} and  quantum spherical $p$-spin glasses  \cite{NieuwRit98,Cugl+01}.

\section{The partition function}\label{sec:Partition}
The quantum statistical partition function for the fixed disorder configuration  $\{J_{ij} \}$ in a basis of coordinate states $|\bm{u}\rangle =| u_1\rangle\ldots | u_N\rangle$ is
\be\label{eq:ZJ}
Z_J=\textrm{Tr}\,\,\textrm{exp}(-\beta \hat{H})=\int d\bm{u}\,\,\langle\bm{u}|\textrm{exp}(-\beta\hat{H})|\bm{u}\rangle,
\ee
where  the Hamiltonian $\hat{H}$ is defined by (\ref{eq:H})  with $u_i$ and $p_i$ replaced by the operators  $\hat{u}_i$ and $\hat{p}_i$. 
In the Matsubara formalism the path integral representation of (\ref{eq:ZJ})  is constructed using the Lie-Trotter product formula \cite{Trotter59,Nelson64}
\be\label{eq:Trot}
\textrm{exp}(-\beta\hat{T}-\beta\hat{V})=\lim_{r\to \infty}\Big\{\textrm{exp}\Big(-\frac{\beta\hat{T}}{r}\Big)\textrm{exp}\Big(-\frac{\beta\hat{V}}{r}\Big)\Big\}^r.
\ee
Insertions of  $\hat{1}=\int d\bm{u}_k|\bm{u}_k\rangle\langle\bm{u}_k|$ and $\hat{1}=\int d\bm{p}_k|\bm{p}_k\rangle\langle\bm{p}_k|
$, together with definitions of imaginary time $\tau_k=k\Delta\tau$ ($k=0,..,r-1$) and time-step  $\Delta\tau=\frac{\hbar\beta}{r}$, 
leads to the following path integral representation of (\ref{eq:ZJ})
\be\label{eq:Z_J}
  Z_J=\int
  {\cal{D}}\bm{u}\,\,\,\textrm{exp}\Big(-\frac{1}{\hbar}{\cal{A}}[\bm{u}]\,\Big),
\ee
where the integration is in the functional sense with a measure defined as
\be
{\cal{D}}\bm{u}=\lim_{r\to \infty}\prod_{k=0}^{r-1}\prod_{i=1}^{N}\sqrt{\frac{mr}{2\pi\hbar^2\beta}}\,\,du_i(\tau_k).
\ee 
The functions $u_i(\tau)$ satisfy the periodicity conditions $u_i(0)=u_i(\hbar\beta)$. 
The Euclidean action reads
\be\label{eq:Aaction}
  {\cal{A}}[\bm{u}]=\int_{0}^{\hbar\beta}
  d\tau\,\Big[\sum_{i=1}^N\frac{m}{2}\Big(\frac{du_i(\tau)}{d\tau}\Big)^2\,+\,V(\bm{u}(\tau))\Big].
\ee
The interaction potential $V(\bm{u}(\tau))$ equals the expression in (\ref{eq:Vu}) with $u_i$ replaced by $u_i(\tau)$.
 
In order to study the equilibrium properties of the model we need to compute the disorder averaged free energy density $f$. The replica trick \cite{MezParVir87} allows us to evaluate this as 
 \be\label{eq:fe}
-\beta f =\frac{1}{N}\overline{\log{Z_J}}=\lim_{n\to 0}\frac{1}{Nn}\log{\overline{(Z_J)^n}},
\ee
where the overline denotes the average over all  realizations of the random interaction.
To end this section we list the expression for $(Z_J)^n$, i.e. the replicated version of (\ref{eq:Z_J})
\be\label{eq:ZJn}
(Z_J)^n=\int\prod_{a=1}^n {\cal{D}}\bm{u}^a\,\,\,\textrm{exp}\Big(-\frac{1}{\hbar}\sum_{a=1}^n {\cal A}[\bm{u}^a]\,\Big).
\ee
The index $a$ numbers the replicas  and $\bm{u}^a=(u_1^a,..,u_N^a)$. 

\section{Effective single-site formulation}\label{sec:SS}
In order to evaluate (\ref{eq:fe}) we first average over all  realizations of the random potential. This is achieved by carrying out the independent Gaussian integrations over the set $\{J_{ij}\}$. 
The result is
\begin{widetext}
\bea\label{eq:ZJn1}
\overline{(Z_J)^n}&=&\int\prod_{a=1}^n {\cal{D}}\bm{u}^a\,\,\,\textrm{exp}\Big[
\sum_{i<j} \Big\{\,\,\frac{J^2}{8\hbar^2 N}\Big(\,\sum_a\int d\tau\,\big[\,u_i^a(\tau)-u_j^a(\tau)\,\big]^2\,\Big)^{2}\,-\,\,
 \frac{g}{\hbar N}\sum_a\int d\tau\,\big[\,u_i^a(\tau)-u_j^a(\tau)\,\big]^4\,\Big\}\,\,\Big]\nonumber\\
&&\qquad\qquad\,\times\,\exp{\Big[-\frac{1}{\hbar}\sum_{ia}\int d\tau\,\,\frac{m}{2}\,\dot{u}_i^a(\tau)^2\Big]}.
\eea
\end{widetext}
The expansions  of the powers in the first line of (\ref{eq:ZJn1}) contain many terms that vanish due to the following   Ansatz. We assume the global $Z_2$-symmetry  to remain unbroken after quantization: $1/N\sum_i u_i^a(\tau)=0$ for all $\tau\in[0,\hbar\beta]$. This means that we do not have to consider terms of the kind $\sum_{ij}\sum_{ab}\int d\tau\int d\tau'\,u_i^a(\tau)^2u_i^b(\tau')u_j^b(\tau')$, which would complicate the formulation considerably. The result after the expansions is
\begin{widetext}
  \bea\label{eq:Zcoupled}
  \overline{(Z_J)^n}&=&\int\prod_{ia}{\cal{D}}u_i^a\,\,\,
  \textrm{exp}\,\Big[\,\,
\sum_{ab}\int d\tau d\tau'\,\Big\{\,\frac{J^2}{4\hbar^2N}\big(\sum_iu_i^a(\tau)u_i^b(\tau')\,\,\big)^2\,+\,\frac{J^2}{8\hbar^2}\sum_iu_i^a(\tau)^2u_i^b(\tau')^2\,\Big\} 
  \nonumber\\
  &&-\frac{1}{\hbar}\sum_{a}\int d\tau\,\,\Big\{\,\sum_i\big(\,\frac{m}{2}\,\dot{u}_i^a(\tau)^2\,+\,g\,u_i^a(\tau)^4\,\big) \,+\,\frac{3g}{ N} \big(\sum_{i} \,u_i^a(\tau)^2 \,\big)^2 \Big\}\,\,\,\Big].
  \eea
\end{widetext}
The sites are decoupled with two sets of Gaussian transformations  after which (\ref{eq:Zcoupled}) becomes
  \bea\label{eq:Zdecoupled}
  \overline{(Z_J)^n}&=&\int {\cal D}\big\{q_{aa}(\tau,\tau)\big\}\,{\cal D}\big\{q_{ab}(\tau,\tau')\big\} \nonumber\\ 
  &&\textrm{exp}\big\{N\big(-\frac{1}{\hbar}{\cal X}[q]\,+\,\textrm{log}\,Z_{\textrm{eff}}\big)\big\},
  \eea
where $Z_{\textrm{eff}}$ defines the effective single-site partition function
  \be\label{eq:Zeff}
  Z_{\textrm{eff}}\,=\,\int\prod_a\,{\cal{D}}u_{a}\,\,\,\textrm{exp}\big(-\frac{1}{\hbar}{\cal{S}}_{\textrm{eff}}\,[u_a]\,\big).
  \ee
The non-fluctuating part in (\ref{eq:Zdecoupled}) is defined as
\bea\label{eq:G}
{\cal X}[q]&=&\frac{J^2}{4\,\hbar}\sum_{ab}\int d\tau d\tau'\,\, q_{ab}(\tau,\tau')^2\nonumber\\
&&-\,3\,g\sum_a\int d\tau\, q_{aa}(\tau,\tau)^2.
\eea
The  effective single-site  action reads
 \be\label{eq:Seff}
  {\cal{S}}_{\textrm{eff}}[u_a]= \frac{1}{2}\sum_{ab}\int
  d\tau d\tau'\,u_a(\tau)
  q_{0,ab}^{-1}(\tau,\tau')u_b(\tau') +{\cal{S}}_{\textrm{int}}[u_a].
\ee
The interaction part ${\cal{S}}_{\textrm{int}}[u_a]$ contains a quartic term non-local in time  and quartic term local in time
\bea\label{eq:Sint}
{\cal{S}}_{\textrm{int}}[u_a]&=&-\frac{J^2}{8\,\hbar}\sum_{ab}\int d\tau d\tau'\,\,u_a(\tau)^2u_b(\tau')^2\nonumber\\
&&+ \,g\sum_a\int d\tau\,\,u_a(\tau)^4.
\eea
The `free'  inverse propagator is
\bea\label{eq:q0}
  q_{0,ab}^{-1}(\tau,\tau')&=&\big\{-m\frac{d^2}{d\tau^2}+12\,g\,q_{aa}(\tau,\tau)\,\big\}\,\delta_{ab}\,\delta(\tau-\tau')\nonumber\\
  &&-\,\frac{J^2}{\hbar}q_{ab}(\tau,\tau')
  \eea
Observe here  that  ${\cal{S}}_{\textrm{eff}}\,[u_a]$ in (\ref{eq:Zeff}) and thus also $Z_{\textrm{eff}}$,   depend functionally on $q_{ab}(\tau,\tau')$. 

The replicated partition function (\ref{eq:Zdecoupled}) can be treated with the saddle-point method. 
At the saddle-points we have  
  \be\label{eq:ZJnfinal}
  \overline{(Z_J)^n}
  \sim\textrm{exp}\Big\{N\big(-\frac{1}{\hbar}{\cal X}[q]\,+\,\textrm{log}\,Z_{\textrm{eff}}\big)\Big\},
  \ee
where the saddle-point fields $q_{ab}(\tau,\tau')$ are the order parameters of the theory. 
The saddle-point equations result in the following functional self-consistency relations for the order parameters
  \be\label{eq:SPE}
  q_{ab}(\tau,\tau')=\langle\, u_a(\tau)\,
  u_b(\tau')\,\rangle\,.
  \ee
The angular brackets $\langle ...\rangle$ denote the quantum thermodynamical average mediated by the effective action (\ref{eq:Seff}). 
The order parameters $q_{ab}(\tau,\tau')$ are the full interacting correlation functions of the single-site theory, from here on called  Matsubara correlations. 
 
The Matsubara correlations are time-translational invariant since we are studying an equilibrium problem. They are also symmetric in  time due to the time-reversal invariance of the action (\ref{eq:Seff}), i.e. we have
  \be\label{eq:SPEsymm}
  q_{ab}(\tau,\tau')\,=\,q_{ab}(\tau-\tau')\,=\,q_{ab}(\tau'-\tau).
  \ee
Furthermore the Matsubara correlations $q_{ab}(\tau-\tau')$ are  $\hbar\beta$ time-periodic.

We should mention  that the first interaction term in (\ref{eq:Sint}) defines a complete square, which could be linearised at the cost of introducing a Gaussian family of systems. However, we have chosen not to do this at this stage. It would lead to  more complicated saddle-point equations for the order parameters when solving the single-site theory perturbatively.  However, we shall linearise this interaction term in the \emph{non}-perturbative treatment.

\section{Perturbation theory}\label{sec:pert}

\subsection{2PI-effective action formalism}
To solve the single-site theory  perturbatively, we need a formalism that expands the path integral (\ref{eq:Zeff}) in terms of 
a further effective (classical) action $\Gamma_{\textrm{eff}}[q]$
\bea\label{eq:Z-Gamma}
Z_{\textrm{eff}}[q]&=&\int\prod_a\,{\cal{D}}u_{a}\,\,\,\textrm{exp}\Big(-\frac{1}{\hbar}{\cal{S}}_{\textrm{eff}}\,[u_a,q]\,\Big)\nonumber\\
&=&\textrm{exp}\Big(-\frac{1}{\hbar}\Gamma_{\textrm{eff}}[q]\,\Big).
\eea
Here we have  explicitly referred to the functional dependences on the Matsubara correlations in ${\cal{S}}_{\textrm{eff}}\,[u_a,q]$ and in $Z_{\textrm{eff}}[q]$. Remember, this dependence is due to the appearance of $q_{ab}(\tau,\tau')$ in the inverse propagator (\ref{eq:q0}). The $q_{ab}(\tau,\tau')$ define the \emph{full interacting} correlations as we saw in the previous section. 
As regards to  a perturbative expansion of the path integral, the most efficient way is to also express $\Gamma_{\textrm{eff}}[q]$ entirely in terms of the full interacting correlations, which was already assumed in the notation in (\ref{eq:Z-Gamma}). 
For this we choose the two-particle irreducible (2PI) effective action approach, developed in field theory \cite{CorJackiwTom74,Klein82}, which is indeed based on an expansion of $\Gamma_{\textrm{eff}}[q]$ in powers of the full interacting correlators $q_{ab}(\tau,\tau')$ and involves only 2PI diagrams. The 2PI nature of the diagrams has the 
additional
 advantage of considerably reducing the number of diagrams that need to be included in the expansion. Also, as the expansion is in terms of the full interacting correlators, the 2PI approach effectively amounts to summing infinite classes of diagrams of a conventional perturbation expansion, thus enabling it to capture effects of a \emph{non}-perturbative nature. Usefulness of the 2PI effective action approach for the study of stochastic dynamical systems was advocated in \cite{Cris+95}. Its application to  the analysis of glassy systems was suggested in \cite{Thes03}.

Before presenting the series expansion of $\Gamma_{\textrm{eff}}[q]$, we discuss the key ingredients of the 2PI effective action approach, in a formulation appropriate for the present problem. 
Following \cite{CorJackiwTom74},
one first  adds  a two-body source term  to the action ${\cal{S}}_{\textrm{eff}}$. This  defines a generating functional 
\bea\label{eq:gen-funct}
 Z_{\textrm{eff}}[K,q]&=&\int\prod_a\,{\cal{D}}u_{a}\,\,\,\textrm{exp}\Big(-\frac{1}{\hbar}\,\Big\{\,{\cal{S}}_{\textrm{eff}}\,[u_a,q] \nonumber\\
&&+\frac{1}{2}\sum_{ab}\int d\tau\,d\tau'\,\,u_a(\tau)K_{ab}(\tau,\tau')u_b(\tau')\,\Big\}\,\Big)\nonumber\\
&\equiv& \textrm{exp}\Big(-\frac{1}{\hbar}W_{\textrm{eff}}[K,q]\Big),
\eea
giving
\be\label{eq:W-eq}
\frac{\delta\, W_{\textrm{eff}}[K,q]}{\delta K_{ab}(\tau,\tau') }=\,\frac{1}{2}q_{ab}(\tau,\tau').
\ee
From (\ref{eq:gen-funct}) and (\ref{eq:G}) follows
\be\label{eq:wq}
\frac{\delta W_{\textrm{eff}}[K,q]}{\delta q_{ab}(\tau,\tau')}+\frac{1}{\hbar}\frac{\delta {\cal X}[q]}{\delta q_{ab}(\tau,\tau')}=0, 
\ee
 We shall need this in the equations of motion for $q_{ab}(\tau,\tau')$, to be constructed next. 
In order to eliminate $K$ in favour of the full interacting correlations $q$, one performs the following Legendre transformation 
\be\label{eq:Legendre}
\Gamma_{\textrm{eff}}[q]=W_{\textrm{eff}}[K,q]-\frac{1}{2}\sum_{ab}\int d\tau\,d\tau'\,q_{ab}(\tau,\tau')K_{ab}(\tau,\tau'),
\ee
giving 
\be\label{eq:Gam-eq}
\frac{\delta \Gamma_{\textrm{eff}}[q]}{\delta q_{ab}(\tau,\tau') }=\frac{\delta W_{\textrm{eff}}[K,q]}{\delta q_{ab}(\tau,\tau')}-\frac{1}{2}K_{ab}(\tau,\tau').
\ee
Then setting the source field $K_{ab}(\tau,\tau')$  to zero and using (\ref{eq:wq}) results in the following `equations of motion' for  $q_{ab}(\tau,\tau')$
\be
\frac{\delta \Gamma_{\textrm{eff}}[q]}{\delta q_{ab}(\tau,\tau') }+\frac{1}{\hbar}\frac{\delta {\cal X}[q]}{\delta q_{ab}(\tau,\tau')}=\,0.\label{eq:Gam-eom}
\ee

Next is to describe the series expansion of $\Gamma_{\textrm{eff}}[q]$, which we present in the standard form as derived in \cite{CorJackiwTom74,Klein82} 
\be\label{eq:G23}
  \Gamma_{\textrm{eff}}[G]=\frac{\hbar}{2}\textrm{Tr}\,q_0^{-1}
  G+\frac{\hbar}{2}\textrm{Tr}\,\textrm{log}\,G^{-1}
+\sum_{p=2}^\infty \Gamma_p[G],
\ee
with Green's function $G_{ab}(\tau,\tau')\equiv q_{ab}(\tau,\tau')/\hbar$. As already mentioned, the  variable $q_{ab}(\tau,\tau')$  defines the full interacting correlation functions and  satisfies the equations of motion (\ref{eq:Gam-eom}). The traces are taken in the functional sense. The first two terms in (\ref{eq:G23}) define what is called the one-loop contribution. The terms denoted by $\Gamma_p[G]$  define the $p$-loop contributions. These are represented by 2PI diagrams  containing $p$ loops. In the next section  we shall  discuss the rules for constructing such diagrams.  A diagram is said to be 2PI if it does not become disconnected upon cutting two lines. The fact that the contributions $\sum_{p=2}^\infty\Gamma_p[G]$  define 2PI diagrams is understood from the following argument. The equations of motion  (\ref{eq:Gam-eom}) with  substitutions of (\ref{eq:G}) and (\ref{eq:G23}) just result in  the Dyson equations 
\be\label{eq:Dyson}
G^{-1}=q_0^{-1}+\,2\,\frac{\delta}{\delta G}\Big(\sum_{p=2}^\infty\Gamma_p[G]+\frac{1}{\hbar}{\cal X}[G]\Big).
\ee
Since the second part of (\ref{eq:Dyson}) defines the proper self-energy, the diagrams of which are known to be one-particle-irreducible, clearly the diagrams of the terms $\Gamma_p[G]$ must be 2PI, (the diagrams of ${\cal X}[G]$ are 2PI). 
In the expansion of 2PI diagrams we shall  consider two orders, the two-loop and the three-loop order. They  are analysed in section \ref{sec:2-loop} and  \ref{sec:3-loop} below. 

One should mention that (\ref{eq:G23}) is only valid for the special case $\langle u_a(\tau)\rangle=0$. If $\langle u_a(\tau)\rangle\equiv {\cal U}_a\ne 0$  a further source term in (\ref{eq:gen-funct}) is needed. This would lead to extra terms   depending on ${\cal U}_a$ in the effective action (\ref{eq:G23}),  and a further equation of motion $\delta \Gamma_{\textrm{eff}}/\delta {\cal U}_a=0$ \cite{CorJackiwTom74}. We shall not concern ourselves with this case since we have assumed global symmetry to remain unbroken, i.e. $\langle u_a(\tau)\rangle =0$ (see section \ref{sec:SS}). 

In the 2PI effective action formalism  the classical form (\ref{eq:Z-Gamma}) results in the following expression for  free energy 
\be\label{eq:fe-perturb}
\beta f^{(2PI)}=\lim_{n \to 0}\frac{1}{n\hbar}\big(\, {\cal X}[q]+\Gamma_{\textrm{eff}}[q]\,\big),
\ee
where we have used  (\ref{eq:fe}) and (\ref{eq:ZJnfinal}).

\subsection{Rules for 2PI diagrams}\label{sec:diagrams}
 The two interaction terms in (\ref{eq:Sint}) determine the following rules for constructing 2PI diagrams and their expressions. Vertices are labelled  by a replica index and a time variable. Two vertices labelled by $(a,\tau)$ and $(b,\tau')$ are connected by a solid line contributing a factor $\hbar G_{ab}(\tau,\tau')=q_{ab}(\tau,\tau')$. The two interaction terms  define two types of vertices. Firstly, there is a  $g$-type vertex which is represented by a dot $\bullet$ and contributes a factor $-g/\hbar$. Secondly, there is a  $J$-type vertex which is represented by a cross $\times$. The $J$-type vertices always come in sets of two. The two are connected by a dashed line $\times ---\times$. A  dashed line contributes a factor $J^2/8\hbar^2$. 
Each diagram gets an extra factor $-\hbar$ due to the prefactor $-1/\hbar$   in (\ref{eq:Zeff}). Then there is a further permutational factor to consider which we write in front of the diagram. 
Next is to collect all factors of a diagram and multiply them. Finally, one  needs to sum over all  replica indices and integrate over all    time variables. 

\begin{figure}[h]
  \includegraphics[width=0.375\textwidth]{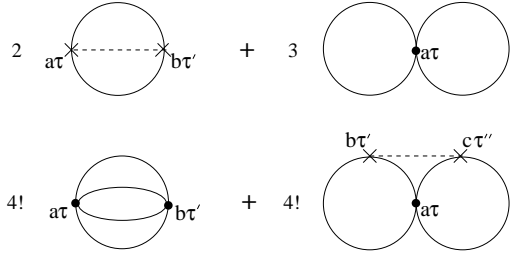}
  \caption{Two-loop and three-loop 2PI diagrams}
  \label{fig:loops}
  \end{figure}

When constructing diagrams of  order $p$, all distinct diagrams containing   $p$ loops  are added together, resulting in the final  expression for  $\Gamma_p[q]$. When counting loops, both solid lines and dashed lines need to be considered. The number of loops in a diagram is equal to the power of $\hbar$ in its expression. A subtlety here is that when counting powers of $\hbar$, one factor $\hbar$ in the contribution of each dashed line  $J^2/8\hbar^2$  must not be taken into account. This is the factor $\hbar$  that originates from the coupling constant $J^2/8\hbar$ of the non-local interaction term  in (\ref{eq:Sint}), which  represents an interaction parameter as a whole.
Finally, it should be noted that  each expansion order $\Gamma_p[q]$ contains  diagrams of ${\cal O}(n)$ in replica and diagrams of ${\cal O}(n^2)$ or higher order. Since $n\to 0$ the latter can be ignored.\\
The 2PI diagrams for two-loop order ($p=2$) and three-loop order ($p=3$) are given in  Fig. \ref{fig:loops}. The prefactors here result from permutations. 

\subsection{Two-loop order}\label{sec:2-loop}
 The two-loop  2PI diagrams  are shown in the first line of Fig. \ref{fig:loops}. 
According to the rules listed in section \ref{sec:diagrams} these diagrams represent the following expressions
\bea\label{eq:G2expr}
\Gamma_2[q]&=&-\frac{J^2}{4\hbar}\sum_{ab}\int d\tau d\tau'\,\,q_{ab}(\tau,\tau')^2\nonumber\\
&&+\,3\,g\sum_a\int d\tau\,\,q_{aa}(\tau,\tau)^2,
\eea
where we have used $q_{ab}(\tau,\tau')= \hbar G_{ab}(\tau,\tau')$. 
 The  equations for the two-loop Matsubara correlations $q_{ab}(\tau,\tau')$ are derived from the equations of motion (\ref{eq:Gam-eom}), resulting in
  \bea\label{eq:SPE-pert2}
  0&=&\frac{1}{2}q_{0,ab}^{-1}(\tau,\tau')-\frac{\hbar}{2}q_{ab}^{-1}(\tau,\tau')
  -\frac{J^2}{2\hbar}q_{ab}(\tau,\tau')\nonumber\\
&&+6g\,\delta_{ab}\delta(\tau-\tau')\,q_{aa}(\tau,\tau).
  \eea
These were solved  in a replica symmetric approximation, and using Fourier transform techniques as described 
in section \ref{sec:per-sol} below.

\subsection{Three-loop order}\label{sec:3-loop}
 The three-loop  2PI diagrams  are shown in the second line of Fig. \ref{fig:loops}.  
These diagrams  represent the following expressions
\bea\label{eq:G23expr}
&&\Gamma_3[q]=-\,\frac{12\,g^2}{\hbar}\sum_{ab}\int d\tau d\tau'\,\,q_{ab}(\tau,\tau')^4\nonumber\\
&&\,\,\,\,+\frac{3J^2g}{\hbar^2}\sum_{abc}\int d\tau d\tau' d\tau''\, q_{ab}(\tau,\tau')^2q_{ac}(\tau,\tau'')^2 
\eea
Again, we derive the  equations for the three-loop Matsubara correlations $q_{ab}(\tau,\tau')$  from the equations of motion (\ref{eq:Gam-eom}),  resulting in
   \bea\label{eq:SPE-pert3}
  0&=&\frac{1}{2}q_{0,ab}^{-1}(\tau,\tau')-\frac{\hbar}{2}q_{ab}^{-1}(\tau,\tau')
  -\frac{J^2}{2\hbar}q_{ab}(\tau,\tau')\nonumber\\
&&+6g\,\delta_{ab}\delta(\tau-\tau')\,q_{aa}(\tau,\tau)
  -\frac{48g^2}{\hbar}q_{ab}(\tau,\tau')^3\nonumber\\
&&+\frac{12J^2g}{\hbar^2}q_{ab}(\tau,\tau')\sum_c \int d\tau''\, q_{ac}(\tau,\tau'')^2.
  \eea

Solutions of the self-consistency equations (\ref{eq:SPE-pert2}) and  (\ref{eq:SPE-pert3}) were attempted in a replica symmetric approximation, and using Fourier transform techniques as described below. First, however let us turn to a description of the two non-perturbative approaches we have looked at, both also in a replica symmetric version.

\section{Non-perturbative theory}\label{sec:non-pert}

\subsection{Replica symmetry}\label{sec:sec-dec}
The non-perturbative theory is constructed in the  replica symmetric (RS) approximation. We mention here that the effects  of replica symmetry breaking  on the low-temperature anomalies have been small for the \emph{semi}-classical treatments \cite{KuHo97,Kuehn03}. 
The RS form of the  Matsubara correlations is
 \begin{subequations}\label{eq:qd-RS}
\begin{align}
q_{aa}(\tau,\tau')&=q_d(\tau,\tau'),\label{eq:qd-RSqd}\\
q_{ab}(\tau,\tau')&=q\,\,(\textrm{for }a\ne b),\label{eq:qd-RSq}
\end{align}
\end{subequations}
independent of the replicas $a,b$. 
 The off-diagonal Matsubara correlation (\ref{eq:qd-RSq}) is assumed  time-independent. The argument is that the replicas are independent and time-translationally invariant, and that the origin of time could be chosen independently for each replica \cite{BrayMoore80}. 

The RS Ansatz allows us to decouple the replicas in the effective single-site action (\ref{eq:Seff}). This concerns 2 terms. 
The first term to decouple is the non-local quartic interaction given in (\ref{eq:Sint}). We linearise this with  a Gaussian variable $\bar{z}$. 
The second term to consider comes from the part of $q_0^{-1}$ in (\ref{eq:q0}) that involves  $q_{ab}(\tau,\tau')$. This is  decoupled   by means of the Gaussian variable $z$. 
After decoupling, the  effective single-site partition function (\ref{eq:Zeff}) becomes
\be\label{eq:ZeffNP}
Z_{\textrm{eff}}=\int{\cal D}z{\cal D}\bar{z}\,\Big\{\int{\cal D}u\,\,\textrm{exp}\Big(-\frac{1}{\hbar}{\cal S}_{\textrm{RS}}[u;z,\bar{z}]\Big)\Big\}^n,
\ee
where ${\cal D}z=dz/\sqrt{2\pi}\,\exp{(-\frac{1}{2}z^2)}$ and the same form for $\bar{z}$. 
The decoupled RS action reads
  \bea\label{eq:Sdec}
  {\cal{S}}_{\textrm{RS}}[u;z,\bar{z}]
&=&\int d\tau\,\,\Big[\,\frac{m}{2}\Big(\frac{du(\tau)}{d\tau}\Big)^2-Jz\sqrt{q}\,u(\tau)\nonumber\\
&&+\frac{1}{2}\Big(12g\,q_d(\tau,\tau)+J\bar{z} \Big)u(\tau)^2+g\,u(\tau)^4\nonumber\\
   &&-\frac{J^2}{2\hbar}\int d\tau'\big(q_d(\tau,\tau')-q\big)u(\tau)u(\tau')\Big]
  \eea
Substituting (\ref{eq:ZeffNP}) in  (\ref{eq:ZJnfinal}) gives for  the free energy (\ref{eq:fe}) 
\bea\label{eq:fe-NP}
\beta f&=&-\int{\cal D}z{\cal D}\bar{z}\,\,\textrm{log}\Big\{\int{\cal D}u\,\,\textrm{exp}\Big(-\frac{1}{\hbar}{\cal S}_{\textrm{RS}}\Big)\Big\}\nonumber\\
&&+\,\,\frac{1}{\hbar}{\cal X}_{\textrm{RS}}[q],
\eea
where the  non-fluctuating part is now defined as
\be\label{eq:GNP}
{\cal X}_{\textrm{RS}}[q]=\frac{J^2}{4\hbar}\int d\tau d\tau'\,\big(q_d(\tau,\tau')^2-q^2 \big)-3\,g\int d\tau\,q_d(\tau,\tau)^2.
\ee
Finally, the RS saddle-point equations (\ref{eq:SPE}) become 
\begin{subequations}\label{eq:SPE-NP}
\begin{align}
  q_d(\tau,\tau')\,=\,&\langle\langle\, u(\tau)\,
  u(\tau')\,\rangle\rangle_{z\bar{z}},\\
  q\,=\,&\langle\langle\, u(\tau)\,\rangle^{2}\rangle_{z\bar{z}},
\end{align}
\end{subequations}
where  $\langle ...\rangle$ denotes an average mediated by the action ${\cal{S}}_{\textrm{RS}}$, and  $\langle ...\rangle_{z,\bar{z}}$ the averages w.r.t. the Gaussians $z$ and $\bar{z}$.
The theory defined by (\ref{eq:Sdec}-\ref{eq:SPE-NP}) is non-local in Matsubara time. There are currently no analytic techniques available to solve the self-consistency problem for the Matsubara correlations in this theory non-perturbatively while keeping the full complexity arising from this fact. Two different methods will be used to deal with this problem, a numerical approach using Quantum Monte Carlo simulations as proposed in \cite{GremRoz98}, and the so-called static approximation due to Bray and Moore \cite{BrayMoore80}.

\subsection{Static approximation}\label{sec:static}
In this approximation  the non-local  term (the third line in (\ref{eq:Sdec})) is approximated by taking for the diagonal Matsubara correlations a simple time-independent trial function 
\be\label{eq:Static-Ansatz}
q_d(\tau,\tau')=q_d,
\ee
which may but need not assumed to be equal to $q_d(\tau,\tau)\equiv q_d(0)$.  Such static approximation  scheme  was  introduced as a variational Ansatz in \cite{BrayMoore80} when studying the spin-glass transition for the SK-model generalised to quantum spins.
After applying (\ref{eq:Static-Ansatz})  we are able to linearise 
the non-local term of (\ref{eq:Sdec}) by means of a further Gaussian transformation, defined by  a Gaussian variable $v$. This results in the following  static  action
  \bea\label{eq:Sstat}
  {\cal{S}}_{\textrm{st}}[u,v;z,\bar{z}]&=&\int d\tau\,\,\Big[\,\,\frac{m}{2}\Big(\frac{du(\tau)}{d\tau}\Big)^2+\,\frac{1}{2}\,v^2\nonumber\\
&&-J\sqrt{C}\,v\,u(\tau)\,+\,d_1(z)\,u(\tau)\nonumber\\
&&+\,\,d_2(\bar{z})\,u(\tau)^2\,+\,g\,u(\tau)^4\,\,\Big],
\eea
with $C=\beta(q_d-q)$. The random parameters $d_1$ and $d_2$  are defined as
\bea\label{eq:d1d2}
  d_1(z)&=&-Jz\sqrt{q}\nonumber\\
  d_2(\bar{z})&=&\frac{1}{2}\big(12\,g\,q_d(0)\,+\,J\,\bar{z}\, \big),
  \eea
with $q_d(0)= q_d(\tau,\tau)$. The free energy (\ref{eq:fe-NP})  becomes
\bea\label{eq:fe-stat}
\beta f&=&
-\int{\cal D}z{\cal D}\bar{z}\,\textrm{log}\Big\{\int{\cal D}u\int\frac{dv}{\sqrt{2\pi/\beta}}\,\,\textrm{exp}\Big(-\frac{1}{\hbar}{\cal S}_{\textrm{st}}\Big)\Big\}\nonumber\\
&&+\,\,\frac{1}{\hbar}{\cal X}_{\textrm{RS}}[q]
\eea
with the non-fluctuating part ${\cal X}_{\textrm{RS}}[q]$ given by (\ref{eq:GNP}).
 
Before presenting the self-consistency relations for the order parameters, let us first take a closer look at the static action (\ref{eq:Sstat}).
The parameters $d_1(z),d_2(\bar{z})$ and $g$ are the coupling constants of a potential energy  for the system defined by the local \emph{quantum} variable $u(\tau)$. 
The (quenched) Gaussians $z$ and $\bar{z}$ imply that we have here a heterogeneous family of such systems.
This results in a broad spectrum of \emph{tunnelling} and  \emph{vibrational excitations}, which we shall discuss in a separate section below.

Interestingly, there is another term in the action namely $-J\sqrt{C}\,v\,u(\tau)$, which is of a different nature. 
The constant $J\sqrt{C}$  defines a coupling constant for the bilinear interaction between the variable $u(\tau)$ and the \emph{classical} (annealed) degree of freedom $v$. The latter is indeed  a classical variable since it has no kinetic term associated with it.
The appearance of this `annealed' degree of freedom is a result of formal mathematical analysis (as are the quenched variables $z$ and $\bar{z}$). Both are `interpreted' and they acquire different meanings due to the different ways in which they appear in the theory. The $z$ and $\bar{z}$ variables are \emph{frozen}, resulting in an ensemble of double-well and single-well potentials, whereas the variable $v$ defines a \emph{dynamical} degree of freedom. The coupling to $v$ is entirely analogous to the coupling to a heat-bath of phonons, as it is postulated in the phenomenological models of glassy low-temperature anomalies \cite{And+72,Phil72,Karp+83,Buch+92}, though the details are of course different.
Whereas phenomenological models \emph{postulate} a coupling of local degrees of freedom to the strain-field of a heat-bath of phonons as an additional ingredient, the coupling to a harmonic classical variable $v$ in the present case \emph{emerges} through the (approximate) mathematical treatment of quantum fluctuations.
Incidentally the presence of a heat-bath like background system could have been inferred directly from the appearance of retarded interactions in (\ref{eq:Seff})-(\ref{eq:q0}) as such retarded interactions are the \emph{usual} hallmark of effective descriptions of systems embedded into larger systems, after intergrating out the degrees of freedom of those larger systems.

Next we evaluate the free energy as a variational estimate  w.r.t. the static Matsubara correlations $q_d(0), q_d$ and $q$. 
The  static formulation (\ref{eq:Sstat}-\ref{eq:fe-stat}) defines a theory local in time. We were able to find numerical solutions for the Matsubara correlations in two different approaches, a  \emph{three-variable} approach in terms of  the variables $q_d(\tau,\tau)\equiv q_d(0),q_d$ and $q$, and a \emph{two-variable} approach in terms of the variables  $ q_d$ and $q$, assuming $q_d(0)=q_d$. 
The variational equations  in the three-variable approach are
\begin{subequations}\label{eq:SPE-3var}
\begin{align}
  q_d(0)\,=\,&\langle\langle u(\tau)^2\rangle\rangle_{z\bar{z}},\\
  \frac{J}{\hbar}q_d\sqrt{C}\,=\,&\langle\langle v\,u(\tau)\rangle\rangle_{z\bar{z}},\\
  q\,=\,&\langle\langle u(\tau)\rangle^2\rangle_{z\bar{z}}.
\end{align}
\end{subequations}
Here $\langle ...\rangle$ denotes an average mediated by the static action (\ref{eq:Sstat}), while  $\langle ...\rangle_{z,\bar{z}} $ denotes the Gaussian averages over $z$ and $\bar{z}$. 
 In the two-variable approach the variational equations are
\begin{subequations}\label{eq:SPE-2var}
\begin{align}
  \Big(\frac{J}{\hbar}\Big)^2C\,=\,&-12\,\frac{g}{\hbar}\,\big(\,\langle\langle u(\tau)^2\rangle\rangle_{z\bar{z}}-q_d\, \big)+\frac{J}{\hbar\sqrt{q}}\langle z\langle u(\tau)\rangle\rangle_{z\bar{z}},\\
  q\,=\,&\langle\langle u(\tau)\rangle^2\rangle_{z\bar{z}}.
\end{align}
\end{subequations}
We will solve the functional self-consistency equations (\ref{eq:SPE-3var}) and (\ref{eq:SPE-2var}) reverting to an operator description, and using truncated Hilbert-spaces as described in section \ref{sec:nonper-sol-static} below.

\subsubsection{Tunnelling and vibrational excitations}\label{sec:DWP-SWP}

The potential energy in the action (\ref{eq:Sstat}) contains an ensemble of single-well and (asymmetric) double-well potentials because of the stochastic nature of the parameters $d_1(z)$ and $d_2(\bar{z})$ as defined in (\ref{eq:d1d2}).
These single-well and double-well potentials are responsible for respectively vibrational excitations and the characteristic tunnelling excitations in the system \cite{Kuehn03}.  
We see that this potential structure arises naturally as a result of \emph{microscopic} interactions defined by the model.
In this context we mention the phenomenological soft potential model \cite{Karp+83,Buch+92} in which one \emph{postulates} the existence of an ensemble of \emph{classical} potentials $V(u)=d_1 u+d_2u^2+gu^4$, providing a semi-classical analysis of its tunnelling- and vibrational states. Whereas that model assumes a \emph{uniform} distribution of the parameters $d_1$ and $d_2$, 
the quantum statistical treatment of the microscopic model as presented here predicts a \emph{Gaussian} distribution of the parameters  $d_1(z)$ and $d_2(\bar{z})$. Furthermore, $d_1(z)$ and $d_2(\bar{z})$ are parameterised by the disorder strength $J$ and the order parameters $q$ and $q_d(0)$. The \emph{collective}  nature of the latter can be seen as the origin of  universality  of the low-temperature physics predicted by the model.

\section{Numerical results}\label{sec:results}

\subsection{Scaling}\label{sec:scaling}
For numerics and representation of results we represent the theory constructed in the previous sections in terms of dimensionless variables and parameters. 
Starting from a microscopic length scale $u_0$ we define the following energy scales
\be\label{eq:dimless-erg}
E_0=\frac{\hbar^2}{mu_0^2},\qquad E_g=g\,u_0^4,\qquad E_J=Ju_0^2,
\ee
where $E_0$ defines the quantum energy scale. 
The dimensionless ratios of the variables are 
\be\label{eq:dimless-var}
\tilde{u}=\frac{u}{u_0},\quad\tilde{q}=\frac{q}{u_0^2}\,.
\ee
Those for the parameters are defined as
\be\label{eq:dimless-par}
\tilde{g}=\frac{E_g}{E_0},\quad\tilde{J}=\frac{E_J}{E_0},\quad\tilde{T}=\frac{k_B T}{E_0}=\tilde{\beta}^{-1}.
\ee
Let us consider the relation between the dimensionless temperature  $\tilde{T}$ and the absolute temperature $T$ for the  simple example of vitreous silica, the amorphous state of $SiO_2$ . Taking a  microscopic length scale  $u_0= 10^{-10}m$ and substituting the values of $\hbar,k_B$ and the mass $m$ of $SiO_2$ in the definitions above,  implies for this case that $\tilde{T}=1$ corresponds approximately to $T=1$ K. 
In this context we shall from here on look at the dimensionless temperature  $\tilde{T}$ as an approximate representation of the  absolute temperature. 

In what follows we shall ignore writing the tildes on the dimensionless variables and parameters. One can show that the scaled form of all equations given in the previous sections is then obtained  by setting  $\hbar=m=1$.

\subsection{Matsubara correlations}\label{sec:matsubara}

\subsubsection{Perturbative solutions at two-loop order}\label{sec:per-sol}
The perturbative Matsubara correlations were computed at two-loop order in the RS approximation. This  requires solving the saddle-point equations (\ref{eq:SPE-pert2}) with $q_{aa}(\tau-\tau')=q_d(\tau-\tau')$ which is an $\hbar\beta$-periodic function, and $q_{a\ne b}(\tau-\tau')=q$. We treated them in a Fourier transformed representation.
 Our convention for the  Fourier transform of $\hbar\beta$-periodic functions $f(\tau)$ is as follows 
\begin{subequations}\label{eq:Fourier-u}
\begin{align}
f(\tau)&=\sum_k\textrm{e}^{i\omega_k(\tau)}\,\,\hat{f}(\omega_k),\\
\hat{f}(\omega_k)&=\frac{1}{\hbar\beta}\int d\tau\,\,\,\textrm{e}^{-i\omega_k(\tau)}\,f(\tau),
\end{align}
\end{subequations}
with Matsubara frequencies  $\omega_k=\frac{2\pi}{\hbar\beta}k$ ($k=0,\pm 1,..$). 
The advantage of this convention is that the dimension of the transformed quantity is equal to its original dimension.
The  Fourier transform of the saddle-point equations (\ref{eq:SPE-pert2}) requires computing the Fourier transform $\hat{q}_{ab}^{-1}(\omega_k)$ of the (functional) inverse kernel $q_{ab}^{-1}(\tau,\tau')$. 
Note that the latter defines an inverse w.r.t. both replica structure and Matsubara-time integration satisfying
\be
\sum_c\int d\tau'' q_{ac}^{-1}(\tau-\tau'')q_{cb}(\tau''-\tau')=\delta_{ab}\delta(\tau-\tau').
\ee
Using Fourier transform relations requires that
\be\label{eq:inverse-1}
\sum_c\hat{q}_{ac}^{-1}(\omega_k)\hat{q}_{cb}(-\omega_k)=\delta_{ab}/(\hbar\beta)^2,
\ee 
for all $k$, i.e. up to a factor $1/(\hbar\beta)^2$ the Fourier transforms $\hat{q}_{ac}^{-1}(\omega_k)$ of the inverse kernel are equal to the corresponding elements of the matrix inverse $\hat{q}(\omega_k)_{ac}^{-1}$ of the $\hat{q}(\omega_k)$ matrix in replica space,
\be\label{eq:Kernel}
\hat{q}_{ab}^{-1}(\omega_k)=\hat{q}(\omega_k)_{ab}^{-1}/(\hbar\beta)^2.
\ee
The RS representation of the matrix elements $\hat{q}(\omega_k)_{ac}^{-1}$ in the $n\to 0$ limit are given by \cite{MezParVir87}
\begin{subequations}\label{eq:RS-Kernel}
\begin{align}
\hat{q}(\omega_k)_{aa}^{-1}=&\frac{1}{\hat{q}_d(\omega_k)-\hat{q}(\omega_k)}-\frac{\hat{q}(\omega_k)}{(\hat{q}_d(\omega_k)-\hat{q}(\omega_k))^2},\\
\hat{q}(\omega_k)_{a\ne b}^{-1}=&-\frac{\hat{q}(\omega_k)}{(\hat{q}_d(\omega_k)-\hat{q}(\omega_k))^2},
\end{align}
\end{subequations}
where $\hat{q}(\omega_k)=\hat{q}\delta_{k,0}$. Using (\ref{eq:RS-Kernel}) and (\ref{eq:Kernel}) in the
Fourier transform of (\ref{eq:SPE-pert2})  leads to the following equations for the two-loop RS Matsubara correlations. 
 First, for $k=0$ we have
\begin{subequations}\label{eq:RS-pert-eq-k0} of  (\ref{eq:SPE-pert2})
\begin{align}
\frac{1}{\hat{q}_d(\omega_0)-\hat{q}}-\frac{\hat{q}}{(\hat{q}_d(\omega_0)-\hat{q})^2}=&-2(\beta J)^2\hat{q}_d(\omega_0)\nonumber\\
&+24g\beta \sum_k\hat{q}_d(\omega_k),\label{eq:RS-pert-eq-k0-1}  \\
-\frac{\hat{q}}{(\hat{q}_d(\omega_0)-\hat{q})^2}=&-2(\beta J)^2\hat{q},\label{eq:RS-pert-eq-k0-2}
\end{align}
\end{subequations}
where (\ref{eq:RS-pert-eq-k0-1}) represents the replica-diagonal and (\ref{eq:RS-pert-eq-k0-2}) the replica off-diagonal case. 
For $k\ne 0$ we only have to consider the replica-diagonal case, giving
\be\label{eq:RS-pert-eq-kne0}
\frac{1}{\hat{q}_d(\omega_k)}=\beta\,\omega_k^2-2(\beta J)^2\hat{q}_d(\omega_k)+24g\beta \sum_k\hat{q}_d(\omega_k)
\ee
In the high-temperature phase where $\hat{q}=0$ the $\hat{q}_d(\omega_k)$ are found numerically from (\ref{eq:RS-pert-eq-k0}) and (\ref{eq:RS-pert-eq-kne0}) by solving a single self-consistency equation for the variable $\hat{z}_1=\sum_k\hat{q}_d(\omega_k)=q_d(0)$, namely
\be
\hat{z}_1=\frac{1}{4\beta J^2}\sum_k\Big( B_k-\sqrt{B_k^2-8J^2} \,\Big),
\ee
in which $B_k=\omega_k^2+24g\hat{z}_1$.
In the low-temperature phase where $\hat{q}\ne 0$, $\hat{z}_1$ can be determined analytically, entailing that the $\hat{q}_d(\omega_k)$ can be expressed in closed form as
\begin{subequations}\label{eq:Exact2loop}
\begin{align}
\beta\big(\hat{q}_d(\omega_0)-\hat{q}\big)=&\frac{1}{J\sqrt{2}},\\
\hat{q}_d(\omega_{k\ne 0})=&\frac{1}{4\beta J^2}\Big(B_k-\sqrt{B_k^2-8J^2}\,\Big),\\
\hat{z}_1=&\frac{\sqrt{2}\,J}{12\,g},
\end{align}
\end{subequations}

 \begin{figure}[h]
  \includegraphics[width=0.45\textwidth]{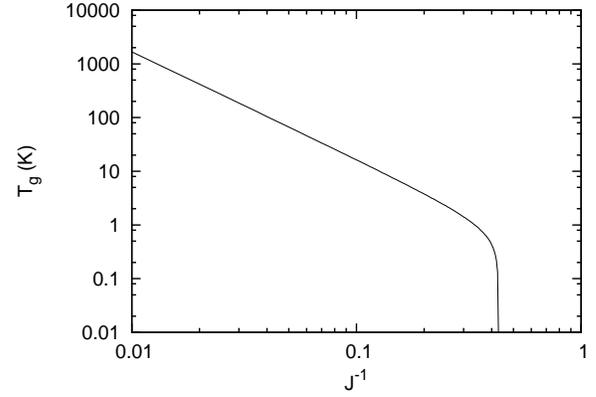}
  \caption{Glass transition temperatures $T_g$ vs. $J^{-1}$ at two-loop perturbation theory for $g=1$. }
  \label{fig:glass-transition}
  \end{figure}
The glass transition temperature $T_g$ as a function of $J$ (at $g=1$) is shown in Fig. \ref{fig:glass-transition}.
Glass transition temperatures for structural glasses are typically in the range between $500\,K$ and $1500\,K$. Clearly this requires $J$ to be large. The majority of our results in the present study were therefore computed for a typical large $J$, $J=50$ giving $T_g\approx 500\,K$.

\begin{figure}[h]
  \includegraphics[width=0.45\textwidth]{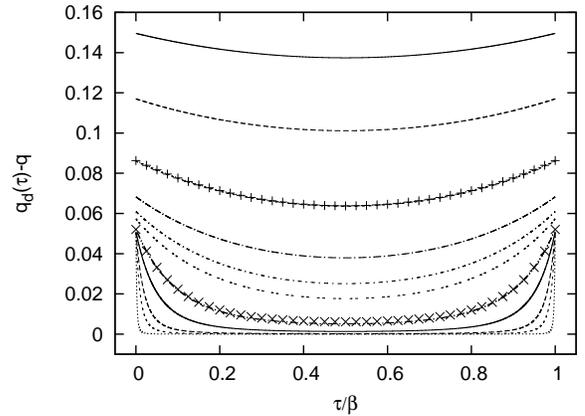}
  \caption{Two-loop perturbative data for the function $q_d(\tau)-q$. Inverse temperatures (in $K^{-1}$) from top to bottom are $\beta=0.1,0.13,0.2,0.3,0.4,0.5,1,2,5,10,50$. Comparison with QMC solutions of the non-perturbative theory for $\beta=0.5$ and $\beta=1$ (marked by $+$ and $\times$).}
  \label{fig:MB1}
  \end{figure}

\begin{figure}[h]
  \includegraphics[width=0.45\textwidth]{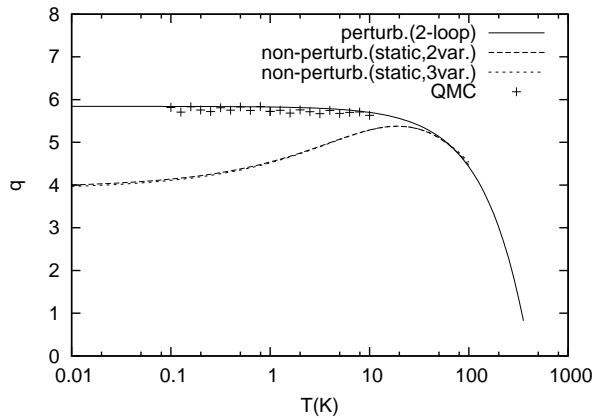}
  \caption{Perturbative and non-perturbative data for the Matsubara correlation $q$  in the low-temperature phase. Note the good agreement of  two-loop perturbative data with  non-perturbative data from Quantum Monte Carlo simulations. Data of non-perturbative static approximations  deviate from these in the low-temperature region.}
  \label{fig:MB2}
  \end{figure}

Next is to discuss the solutions  of (\ref{eq:Exact2loop}) for the Matsubara correlations. The $q_d(\tau)$ were computed from a numerical inverse Fourier transformation of $\hat{q}_d(\omega_k)$. In Fig. \ref{fig:MB1} we plot the results for $q_d(\tau)-q\,$ for a number of low temperatures. A selection of them is compared with the results of  Quantum Monte Carlo (QMC) simulations  of the \emph{non}-perturbative theory,
which will be discussed in a subsection below.
The solutions for the Matsubara correlation  $q$  are plotted in  Fig. \ref{fig:MB2} in the low-temperature phase. 
From both Fig. \ref{fig:MB1} and \ref{fig:MB2} we  conclude that the two-loop results  are in reasonably good agreement with the QMC results. 

The Fourier-representation of the self-consistency equations for the Matsubara correlations  in the low-temperature phase at three-loop order is given in the appendix. The coupling between the Fourier modes at three-loop order contains a truly functional element via the function $\hat{z}_3(\omega_k)$. Although we succeeded in simplifying the problem to solving a set of only four coupled transcendental equations for the variables $\hat{z}_1,\hat{z}_2,\hat{q}_d(\omega_0)$ and $q$, we have so far been unable to solve them. In fact we suspect that physically acceptable solutions may not exist at three-loop order and an expansion to higher loop order might be necessary.

\subsubsection{Non-perturbative Quantum Monte Carlo simulations}\label{sec:nonper-solQMC} 
The  Matsubara correlations $q_d(\tau)$ and $q$ of the non-perturbative RS theory defined in section \ref{sec:sec-dec} above, were evaluated with  Quantum Monte Carlo  simulations. 
This involved solving the functional self-consistent relations (\ref{eq:SPE-NP}).  We used iterative QMC-techniques along the  lines of \cite{GremRoz98}, starting with   a set of initial values of $q_d(\tau)$ and $q$ to be used as input for the action (\ref{eq:Sdec}), after which they were updated in a path integral Monte Carlo algorithm. This procedure was repeated 10 times, resulting in reasonably good convergence of the Matsubara correlations. As regards to the algorithm, 
an update contained $10^5$ Monte Carlo sweeps (taking data every 10th sweep), $\,5\cdot 10^4$ equilibration sweeps and $5\cdot 10^3$ Gaussian $z,\bar{z}$ samples. The imaginary time axis was discretised into 40 time slices. The results for $q_d(\tau)-q\,$ at a selected  number of temperatures are plotted in Fig. \ref{fig:MB1}. The results for the off-diagonal Matsubara correlation $q$ are plotted separately in Fig. \ref{fig:MB2}.
 As mentioned previously, they were found to be in good agreement with the solutions of the two-loop perturbative theory given in (\ref{eq:Exact2loop}). The  conclusion from the simulations is that they confirm the validity of the perturbative \emph{two}-loop results for the Matsubara correlations.

\subsubsection{Non-perturbative static solutions}\label{sec:nonper-sol-static} 
The Matsubara correlations of the static approximation  treated in section \ref{sec:static}, were computed numerically. 
This involved solving the functional self-consistency relations (\ref{eq:SPE-3var}) and (\ref{eq:SPE-2var}). They were  solved in the  operator representation, for which  the  required  Hamiltonian $\hat{H}_{\textrm{st}}(\hat{p},\hat{u},v;z,\bar{z})$ is reconstructed from  the action (\ref{eq:Sstat}).  
Path integrals are then re-expressed in terms of traces over a suitable truncated Hilbert space. We used a bases of harmonic oscillator eigenstates. The Gaussian integrals integrals were computed with Gauss-Legendre quadratures.

The results for the replica off-diagonal Matsubara correlations $q$  are plotted in Fig. \ref{fig:MB2} over a large temperature range. We believe the differences with the   two-loop perturbative  and QMC  results seen here to be an artifact of the static approximation, which after all does not represent the true self-consistent solutions of the saddle-point equations (\ref{eq:SPE-NP}).

\begin{figure}[h]
  \includegraphics[width=0.4\textwidth]{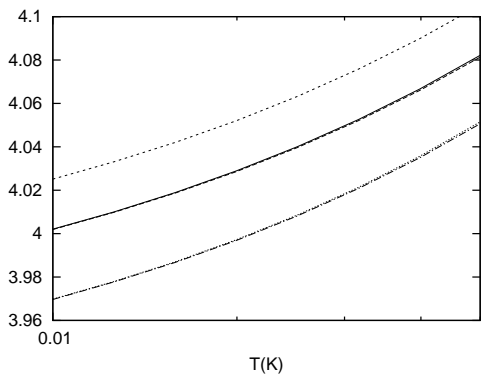}
  \caption{The upper line shows $q_d(0)$ in the three-variable approach. The lowest pair of lines correspond to $q_d$ and $q$ in the three-variable approach, whereas the pair of lines in the middle correspond to $q_d$ and $q$ in the two-variable approach.}
  \label{fig:qdq3var}
  \end{figure}

Differences between the results of the two-variable and three-variable approach are not visible on the scale used in Fig. \ref{fig:MB2}.  They are plotted separately in Fig. \ref{fig:qdq3var} for a selected range of  low temperatures. The combination of order parameters $C=\beta(q_d-q)$, which can be seen as a susceptibility-like variable,  was found to be  approximately equal for both, the  two-variable and the three-variable approach, with nearly constant numerical value $C\approx 17\cdot 10^{-3}$ for $0<T<100\,K$. Since this is very small, the values of $q_d$ and  $q$ barely differ  at low temperatures, as is indicated in Fig. \ref{fig:qdq3var}. 
On the other hand, the two-variable and three-variable approaches do give rise to different values for the $(q_d,q)$ pairs: Introduction of a third variable $q_d(0)$ in the three-variable approach leads to a depression of the values of $q_d$ and $q$ relative to those in the two-variable approach, whereas $q_d(0)$ in the three-variable approach turns out to be larger than $q_d$ and $q$ within the two-variable solution.

\subsection{Thermodynamics}\label{sec:thermodynamics}

\subsubsection{Perturbative specific heat}\label{sec:per-sh}
To obtain an expression for the perturbative free energy we substitute  (\ref{eq:G}) and  (\ref{eq:G23})  in (\ref{eq:fe-perturb}). 
We consider Fourier transforms as defined in (\ref{eq:Fourier-u}) and the scaling introduced in section \ref{sec:scaling}.
The trace $\frac{1}{2}\textrm{Tr}\,\hat{q}_0^{-1}
  \hat{q}$  diverges when substituting  $\hat{q}_{0}^{-1}$ from its definition (\ref{eq:q0}). Instead we substitute  the expression for $\hat{q}_{0}^{-1}$  determined by the saddle-point equations (\ref{eq:SPE-pert2}) or (\ref{eq:SPE-pert3}). The result  is then finite up to an irrelevant infinite constant $\frac{1}{2}\textrm{Tr}\,\hat{q}^{-1}\hat{q}$. 
The trace  $\frac{1}{2}\textrm{Tr}\,\textrm{log}\,\hat{q}^{-1}$  can be shown to have the following RS representation \cite{MezParVir87}
\bea\label{eq:RS-Trlog}
\frac{1}{2}\textrm{Tr}\,\log\,\hat{q}^{-1}&=&-\frac{n}{2}\sum_k\Big\{ \frac{\hat{q}(\omega_k)}{\hat{q}_d(\omega_k)-\hat{q}(\omega_k)}\nonumber\\
&&+\,\log\big( \hat{q}_d(\omega_k)-\hat{q}(\omega_k) \big) \Big\},
\eea 
where $n$ defines the number of replicas. 
Observe that when substituting $\hat{q}(\omega_k)=\hat{q}\delta_{k,0}$ the second part of (\ref{eq:RS-Trlog}) contains the divergent sum $-\frac{1}{2}\sum_{k\ne 0}\textrm{log}\,\hat{q}_d(\omega_k)$.
To deal with this we first   substitute  $\hat{q}_d(\omega_k)$ from (\ref{eq:RS-pert-eq-kne0}) or (\ref{eq:RS-pert-eq-kne0-3loop}).
From the result we isolate a divergent contribution of the following form  $\frac{1}{2}\sum_k\log\big\{\beta(\omega_k^2+24g\hat{z}_1) \big\}$ with $\hat{z}_1=\sum_k\hat{q}_d(\omega_k)$. 
This term, as part of (\ref{eq:RS-Trlog}) and $\Gamma_{\textrm{eff}}$ in (\ref{eq:G23}), should be exponentiated according to (\ref{eq:Z-Gamma}), defining the partition function of a simple harmonic oscillator with frequency $\omega_0=\sqrt{24g\hat{z}_1}$. Consequently, we may replace $\frac{1}{2}\sum_k\log\big\{\beta(\omega_k^2+\omega_0^2) \big\}$ by ($\beta$ times) the free energy of a simple harmonic oscillator, giving $\,\log\sinh(\frac{1}{2}\beta\omega_0)$ which is now finite.

\begin{figure}[h]
  \includegraphics[width=0.45\textwidth]{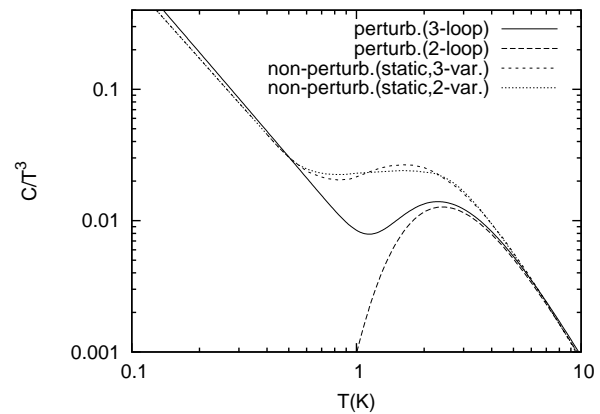}
  \caption{Approx. linear specific heat $C\sim T^{1.02}\,(T<0.8\,K)$ and $C\sim T^{1.1}\,(T<0.5\,K)$ for resp. perturbative (three-loop) and non-perturbative (static) theory and a Bose-peak at higher temperatures. The two-loop perturbative theory described only the vibrational excitations. }
  \label{fig:SHA}
  \end{figure}

To compute the RS free energy numerically we have only   the two-loop data from (\ref{eq:Exact2loop}) at our disposal. The specific heat computed from the two-loop free energy did \emph{not} result in a glassy low-temperature anomaly. Only  vibrational excitations of the system featured here (see Fig.  \ref{fig:SHA}). 
On the other hand, when evaluating the free energy at three-loop order, a specific heat exhibiting the characteristic glassy low-temperature anomaly was obtained ($C\sim T^{1.02}$ for $T<0.8\,K$), though we had to use two-loop results for the Matsubara correlations in those expressions (as three-loop results are so far unavailable). The good agreement between two-loop Matsubara correlations and QMC results is thought to provide a reasonable justification for this approach. 

\subsubsection{Non-perturbative specific heat}\label{sec:nonper-sh}
The non-perturbative thermodynamics was evaluated in the \emph{static} approximation.
For this we used the free energy expression (\ref{eq:fe-stat}) and  the numerical results determined from   (\ref{eq:SPE-3var}) and (\ref{eq:SPE-2var}). This indeed reproduced the characteristic glassy low-temperature specific heat anomaly for both the two-variable and three-variable approach (see Fig.  \ref{fig:SHA}). Again the low-temperature specific heat showed an approximately linear (in fact super-linear) temperature dependence $C\sim T^{1.1}$ for $T<0.5\,K$, in reasonable agreement also with experimental data  \cite{ZellPohl71,Gil+93}. These results should also be compared 
with, and are indeed comparable to those of the translationally invariant model investigated in \cite{Kuehn03}. 
 We found little difference in the results for the  two-variable  and  three-variable approach at higher temperatures, as can be seen in  Fig.  \ref{fig:SHA}. 
Differences between the three-loop perturbative specific heat and the non-perturbative `static' specific heat are restricted to the $0.5-5\, K$ temperature region. 
This could be at least partly because of the strong temperature dependence of the  `static' Matsubara correlations in this region, as displayed  in Fig. \ref{fig:MB2} for the order parameter $q$.

Finally we remark that both the properties of the Bose-peak at intermediate temperatures and the universal tunneling regime at low temperatures are governed by one and the same set of system parameters appearing in (\ref{eq:Sstat}). No separate sets of assumptions were introduced to describe these two temperature regimes.

\section{Conclusions}\label{sec:conclusions}
In summary, we have provided a fully quantum statistical analysis
of a microscopic model of a glass, respecting global translation
invariance. Until now such analysis was available only at a 
semi-classical level. We formulated an effective theory in terms of
single-site path integrals and constructed perturbative and 
non-perturbative solutions of a set of self-consistency equations
describing the system. Both resulted in an approximately linear 
specific heat at low temperatures, in good agreement with experiment.

The perturbation theory was formulated in terms
of two-particle irreducible diagrams at two-loop and three-loop
order for the effective action. As solutions of the self-consistency
equations at three-loop order remained unavaible, we resorted to 
investigating the reliability of our two-loop results using
Quantum Monte Carlo simulations. We found surprisingly good agreement 
of the Matsubara correlations obtained perturbatively and via QMC
simulations.

Within a non-perturbative static approximation we 
obtain a description in terms of a glassy potential energy landscape 
containing an  ensemble of effective single-well and double-well 
potentials, much as in the soft-potential model \cite{Karp+83,Buch+92} and in the 
semi-classical approach \cite{Kuehn03}. Interestingly there is an important 
difference, namely the emergence of a coupling to an additional 
classical variable in a manner reminiscent of a coupling of local 
excitations to a heat bath as postulated within phenomenological models
\cite{Phil81}.

It would be interesting to carry the perturbative approach to higher
loop order, or in fact attempt summations of infinite classes of 
2PI diagrams. On another front, effects of replica symmetry breaking
have not yet been looked at and are worth investigating (though in
a semi-classical approach RSB effects were found to be weak \cite{Kuehn03}).

One of the motivations for the present investigation was to understand 
a phase transition observed in ultra-cold glasses more than a decade 
ago \cite{str+98}, which has so-far not found an explanation.
Regrettably, the present study has not produced any progress in that
particular direction. It might well be the case that an expansion of 
the present investigation in {\em both\/} directions mentioned above 
--- including effects of replica symmetry breaking and inclusion of 
diagrams up to arbitrarily high loop order --- would be required to 
reveal pertinent signatures  of that phase transition.

\section{Appendix}
Here we present the Fourier transformed RS representation of the \emph{three}-loop perturbative equations (\ref{eq:SPE-pert3}) in the low-temperature phase. 
Considering the scaling introduced in section \ref{sec:scaling}  and using (\ref{eq:Kernel}) and (\ref{eq:RS-Kernel}), the Fourier transformation leads to the following set of equations.
First, for $k=0$ we have 
\begin{subequations}\label{eq:RS-pert-eq-k0-3loop}
\begin{align}
\frac{1}{\hat{q}_d(\omega_0)-\hat{q}}-\frac{\hat{q}}{(\hat{q}_d(\omega_0)-\hat{q})^2}=&-2(\beta J)^2\hat{q}_d(\omega_0)+24g\beta \hat{z}_1 \nonumber \\
&-96(\beta g)^2\,\hat{z}_3(\omega_0)\nonumber\\
&+24(\beta J)^2g\,\hat{q}_d(\omega_0)\,\beta(\hat{z}_2-\hat{q}^2  )
,\label{eq:RS-pert-eq-k0-1-3loop}\\
-\frac{\hat{q}}{(\hat{q}_d(\omega_0)-\hat{q})^2}=&-2(\beta J)^2\hat{q}\nonumber\\
&-96(\beta g)^2\,\hat{q}^3\nonumber\\
&+24(\beta J)^2g\,\hat{q}\,\beta(\hat{z}_2-\hat{q}^2  )
,\label{eq:RS-pert-eq-k0-2-3loop}
\end{align}
\end{subequations}
where (\ref{eq:RS-pert-eq-k0-1-3loop}) represents the replica-diagonal and (\ref{eq:RS-pert-eq-k0-2-3loop}) the replica off-diagonal case. 
The last two terms in these equations define the \emph{three}-loop extensions of the two-loop equations (\ref{eq:RS-pert-eq-k0}). Here we used the following definitions
\begin{subequations}\label{eq:z-definitions}
\begin{align}
\hat{z}_2=&\sum_k\hat{q}_d(\omega_k)^2,\\
\hat{z}_3(\omega_k)=&\sum_{lm}\hat{q}_d(\omega_l)\,\hat{q}_d(\omega_m)\,\hat{q}_d(\omega_k-\omega_l-\omega_m).
\end{align}
\end{subequations}
For $k\ne 0$ we only have to consider the replica-diagonal case, giving
\bea\label{eq:RS-pert-eq-kne0-3loop}
\frac{1}{\hat{q}_d(\omega_k)}&=&\beta\,\omega_k^2-2(\beta J)^2\hat{q}_d(\omega_k)+24g\beta \hat{z}_1\nonumber\\
&&-96(\beta g)^2\,\hat{z}_3(\omega_k)\nonumber\\
&&+24(\beta J)^2g\,\hat{q}_d(\omega_k)\,\beta(\hat{z}_2-\hat{q}^2  ),
\eea
where the last two terms again represent the three-loop extensions of the two-loop equations.
The solutions of (\ref{eq:RS-pert-eq-k0-3loop}) can no more  be expressed in  analytic form as was  the case for the two-loop perturbative equations (\ref{eq:Exact2loop}).

\bibliography{MyBib}

\end{document}